\documentclass[sigconf]{acmart}

\AtBeginDocument{%
  \providecommand\BibTeX{{%
    \normalfont B\kern-0.5em{\scshape i\kern-0.25em b}\kern-0.8em\TeX}}}

\setcopyright{acmcopyright}
\copyrightyear{2021}
\acmYear{2021}
\usepackage{multirow}
\begin{document}

\title{"Deep Cut": An all-in-one Geometric Algorithm 
for Unconstrained Cut, Tear and Drill of Soft-bodies 
in Mobile VR}


 \author{Manos Kamarianakis}
 \orcid{0000-0001-6577-0354}
 \authornote{Corresponding Author, \url{kamarianakis@uoc.gr}}
 \affiliation{%
   \institution{University of Crete, ORamaVR}
   \city{Heraklion}
   \country{Greece}
   }
 \author{Nick Lydatakis}
 \orcid{0000-0001-8159-9956}
 \affiliation{%
   \institution{University of Crete, ICS-FORTH, ORamaVR}
   \city{Heraklion}
   \country{Greece}
   }
 \author{Antonis Protopsaltis}
 \orcid{0000-0002-5670-1151}
 \affiliation{%
   \institution{University of Western Macedonia, ORamaVR}
   \city{Veria}
   \country{Greece}
   }

 \author{John Petropoulos}
 \orcid{0000-0001-5373-8760}
 \author{Michail Tamiolakis}
 \orcid{0000-0002-9393-3138}
 \affiliation{%
   \institution{University of Crete, ORamaVR}
   \city{Heraklion}
   \country{Greece}
   }

 \author{Paul Zikas}
 \orcid{0000-0003-2422-1169}
 \affiliation{%
   \institution{University of Crete, ICS-FORTH, ORamaVR}
   \city{Heraklion}
   \country{Greece}
   }

 \author{George Papagiannakis}
 \orcid{0000-0002-2977-9850}
 \affiliation{%
   \institution{University of Crete, ICS-FORTH, ORamaVR}
   \city{Heraklion}
   \country{Greece}
   }

 \renewcommand{\shortauthors}{Kamarianakis, Protopsaltis, Lydatakis et al.}

\begin{abstract}
In this work, we present an integrated geometric framework: "deep-cut" that 
enables for the first time a  user to geometrically and algorithmically cut, tear and drill the surface of a skinned model without prior constraints,
layered on top of a custom soft body mesh deformation algorithm. 
Both layered algorithms in this framework yield real-time results and are amenable for mobile Virtual Reality, in order to be utilized in a variety 
of interactive application scenarios. Our framework dramatically improves real-time user experience and task performance in VR, without pre-calculated or artificially designed cuts, tears, drills or surface deformations via predefined rigged animations, which is the current state-of-the-art in mobile VR. Thus our framework improves user experience on one hand, on the other hand saves both time and costs from expensive, manual, labour-intensive design pre-calculation stages.
\end{abstract}



\begin{teaserfigure}
  \includegraphics[width=\textwidth]{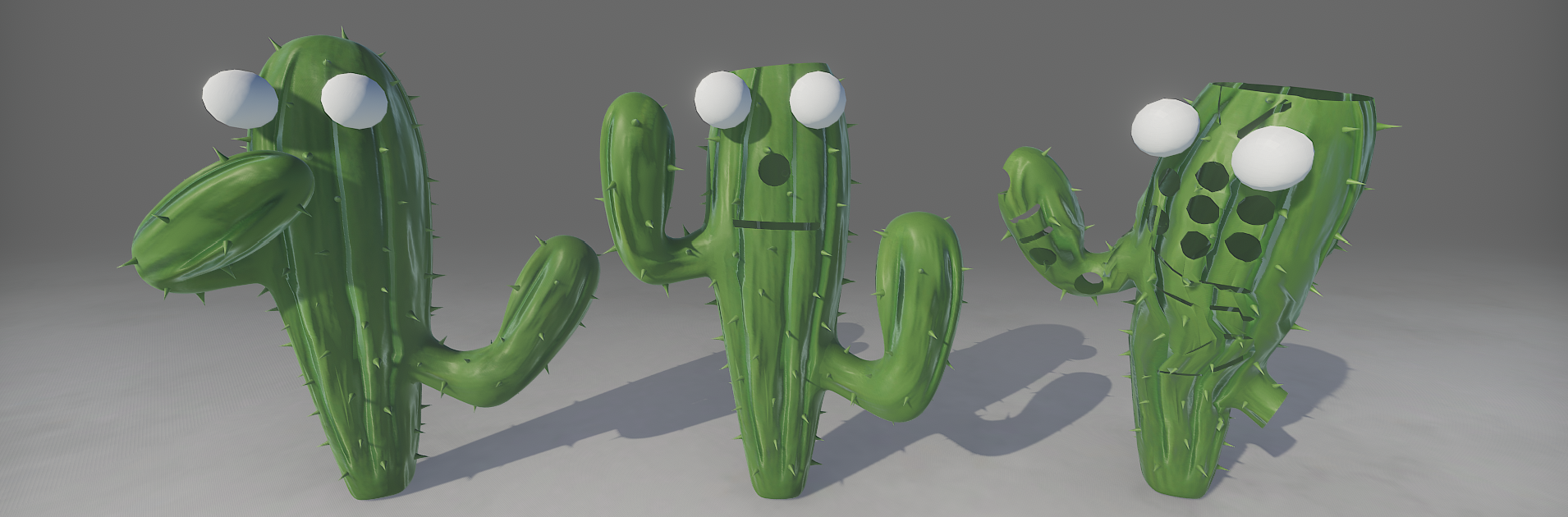}
  \caption{A linear-blend skinned 3D model (left), current state-of-the-art of predefined cut, tear and drill (center), our 'deep-cut' framework for unconstrained, arbitrary cuts, tears, drills with soft-body deformations in mobile VR (right)}
  \Description{Three soft-body cacti are cut, teared, drilled and
  in some cases deformed.}
  \label{fig:teaser}
\end{teaserfigure}

\maketitle

\section{Introduction}

Since their inception, rigged animated models \cite{magnenat1988joint} 
have become a major research
topic in real-time computer graphics. 
Experts have been experimenting with 
various animation and deformation techniques, pushing the boundaries of realism and real-time performance. 
As the industry of Virtual, Augmented Reality (VR, AR) 
rapidly grows, increasingly more complicated and optimized methods are being
developed. 
An example of such tools in computer graphics involve the ability to perform cuts, tears and drills on the surface of a skinned model
\cite{Bruyns:2002jc,Kamarianakis_Papagiannakis_2021}. 
Such algorithms are utilized with the aim to increase user immersion and be used as sub-modules of even more complex operations. However, their scale-up for the extreme real-time conditions of virtual reality environments and the specifically mobile, all-in-one un-tethered VR computing environments, remains an active field of research.

The need to interact in a shared virtual world with other participants 
soon led to the need of more realistic deformation techniques and interaction paradigms. Certain rigid objects, 
e.g., a sponge, are expected to change in real world when external forces are applied on them. 
It feels natural to try and replicate this 
behavior in VR too 
\cite{terzopoulos1987elastically,macklin2014unified,mages_poster}. 
One way to accomplish this 
is via the so-called 
soft-body mesh deformation \cite{mages3}, 
an algorithm that essentially 
dictates how the vertices of a mesh should affect one another when 
an external force is applied somewhere on the surface of the model.  

Performing interactive cuts on a model is not something new; current 
bibliography describes a great number of how one may accomplish this. 
However, most techniques cannot be applied in applications demanding 
a high frame rates, such as computer games or demanding VR applications. 
Moreover, if cuts are indeed implemented in such an application 
they are almost always constrained: camera, model or user degrees of freedom i.e. the user cannot freely cut anywhere
on the model; a set of predefined cuts and their animations are 
usually created by VR designers or artists where each one of them is played/triggered
by the user's specific and constrained actions.

In this work, we propose a framework that allows the user to
perform on a rigged soft-body mesh, actions such as cuts, tears and 
drills on the model's skin surface, without any constraints.
Our algorithms are based on pure geometric operations on the surface mesh, 
and therefore are amenable to yield real-time results, 
even in low-spec devices 
such as mobile VR Head Mounted Displays (HMDs). 
An offline pre-processing of the model's mesh allows the creation of suitable 
data structures that store both vertex neighbours (used in cut/tear/drill) 
and a custom particles grid with weights assignment (used in soft-body deformations). 
This allows the effective, latency-free application of multiple 
consecutive  actions as the data structures
only need to be incrementally updated and not re-created from scratch. 
Our methods can be implemented in modern game engines such as Unity and 
Unreal Engine; convincing results are illustrated in the video 
accompanying this work.

The significance of our work lies on the fact that in the current 
state-of-the-art, cuts, tears and drills of a rigged 3D model in VR are 
predefined and require a VR designer to manually manipulate and rig the model 
for several labour-intensive hours, depending on the model complexity, 
to achieve a single operation of the above. 
Using our methods, the VR designer is removed from the loop, while similar results are obtained at a fraction of time (actually milliseconds). Moreover , in cases where a simple change on a cut, tear, or drill is needed, our proposed algorithms may be applied continuously and incrementally until a satisfying result is achieved, saving a lot of processing time.

Our work is organized as follows: first we provide an outline of our 
methodology (Section~\ref{sec:methodology}), 
we then present our results (Section~\ref{sec:results}) followed by our
conclusions and future work (Section~\ref{sec:conclusions_future_work}).

\section{Our Methodology}\label{sec:methodology}

The key components of our proposed method are the Cut/Tear/Drill 
algorithms, described in 
Section~\ref{sub:CTD}, and 
the Soft Mesh Deformation algorithms, 
described in Section~\ref{sub:SMD}. 
To effectively combine these methods and achieve real-time performance, 
we used a variety of optimization techniques and tools; you may
see Section~\ref{sub:tools} for more details.

\subsection{Algorithms for Cut, Tear and Drill}\label{sub:CTD}

The idea of performing cuts on a model's skin has been a 
highly researched topic \cite{Bruyns:2002jc}. 
We may distinguish two main categories on how this may be achieved. 

The first approach is via Finite Element Methods (FEM), which is 
mainly used to slice parts of concrete objects, using tetrahedral 
meshes \cite{wu_efficient_2013}.  
This is a process requiring heavy preprocessing and near real-time 
results can only be achieved using powerful CPU/GPU machines. 

The second approach utilizes elementary geometric 
subpredicates (e.g., intersection of a face with a given plane) 
on the model mesh. 
Such techniques require far less process power and therefore are 
suitable for VR environments, especially collaborative ones.

\begin{figure}[tbp]
  \centering
  \includegraphics[width=0.45\textwidth]{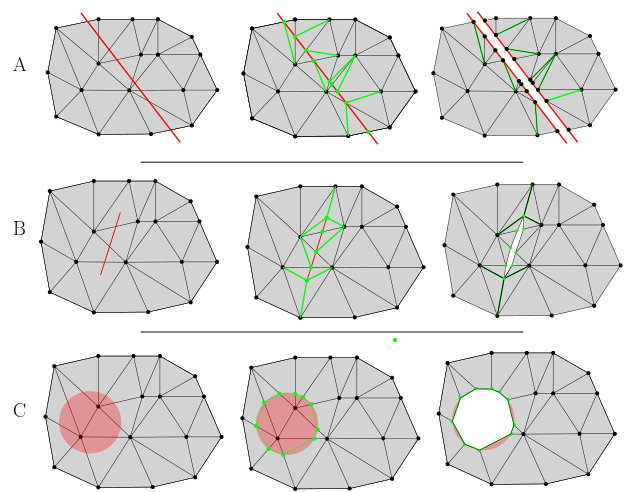}
  \Description{Cut, Tear and Drill in 2D Mesh.}
  \caption{Performing (A) Cut, (B) Tear and (C) Drill on a 
  triangular mesh. 
  The red line represents a plane in (A) and (B). For the Drill 
  case, we simply project the 3D-mesh to a plane and therefore 
  reduce the face versus drill predicate to a simpler 2D-triangle 
  versus circle.}
  \label{fig:ctd}
\end{figure}

Below we provide an outline of how we perform cuts, tears and 
drills, following the second approach (see Figure~\ref{fig:ctd}). 
These algorithms are explained in detail in 
\cite{kamarianakis2020deform} and 
\cite{Kamarianakis_Papagiannakis_2021}.

\textbf{Cutting Algorithm.} The user inputs a plane that intersects 
the model. The algorithm detects the intersection points of the 
mesh with the plane, re-triangulates the mesh and splits the original 
mesh into multiple models (usually two), that correspond to the 
submodels having vertices on each side of the plane only. 

\textbf{Tearing Algorithm.} The user inputs the tearing tool's 
(\emph{scalpel's}) starting and ending positions. 
Let $\Pi$ be the plane defined by the tool's end position and the 
initial intersection point of the model with the scalpel. 
We evaluate the intersection points of $\Pi$ with the model 
that lie between the scalpel's starting and ending position, 
and correspond them to the tear. 
These points are duplicated, re-triangulated 
and moved away from the plane by a fraction of its normal, forming 
a visible tear. 

\textbf{Drilling Algorithm.} The user inputs the radius and the 
end points of the axis  of a cyclical drill that intersects the model. 
The model’s mesh is up-projected to a plane $\Pi$, perpendicular to the drill’s axis, and the mesh vertices intersected by the drill are detected.
After removing all parts of the mesh ``inside'' the drill, we 
re-triangulate the mesh on $\Pi$ and down-project it to the original 
model. 

After performing any of these actions, we assign weights to the 
newly introduced intersection points, using the \emph{update weight} 
function described in \cite{Kamarianakis_Papagiannakis_2021}. 
This allows the final model, or submodels in case of cut, to be  
(re)animated in a artifact-free way.

Details on the preprocessing of the model and the 
data-structures used to perform either of the methods are 
found in Section~\ref{sub:tools}.

\subsection{Soft Body Meshes}\label{sub:SMD}

There are various methods 
to implement soft body mesh deformation for skinned models 
\cite{nealen2006physically}. 
In most methods, the vertices of the mesh are
clustered into groups, called \emph{particles}.
A vertex can belong to multiple particles and a particle
is centered at the average position of all 
vertices it contains. 
Deciding a suitable clustering is not an easy task; 
it depends on the topology of the model 
and the desired deformation behavior the user wants 
to achieve.

\begin{figure}[tbp]
  \centering
  \includegraphics[width=0.45\textwidth]{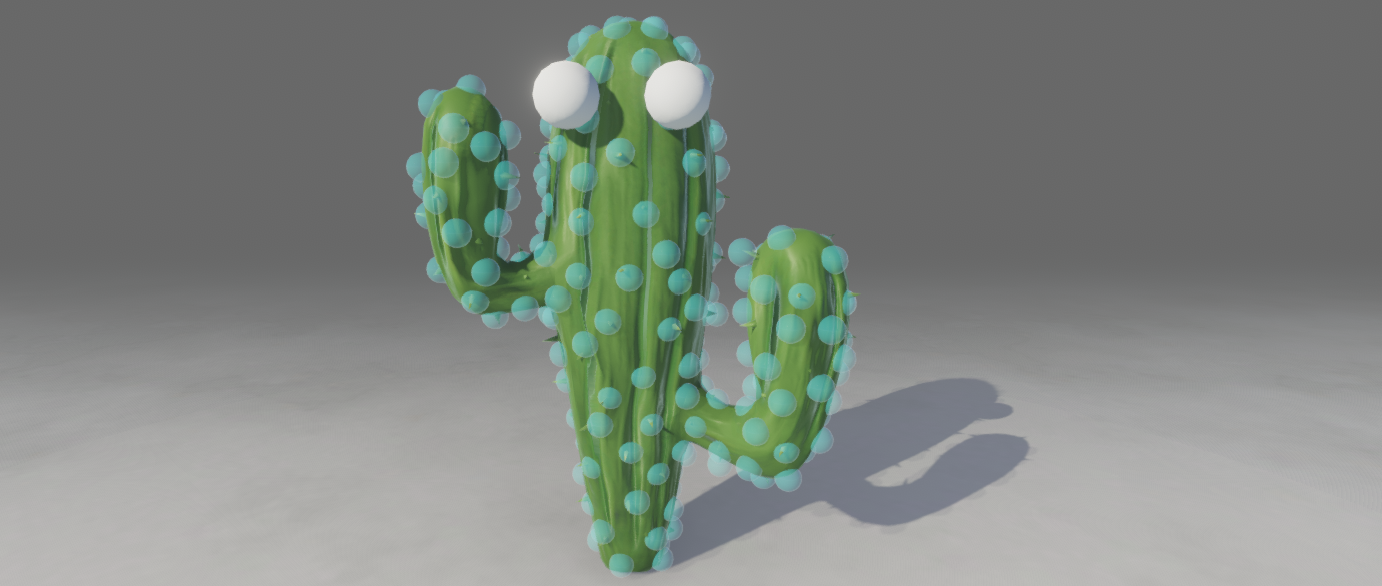}
  \Description{A cactus model is clustered in particles.}
  \caption{The 224 particles of the cactus model used 
  for the soft body deformation.}
  \label{fig:soft_body}
\end{figure}

In our method, the clustering was determined by the 
following property: every particle may contain vertices within 
a range of 0.8 units, using the euclidean distance. 
This resulted in a total of 224 particles (see 
Figure~\ref{fig:soft_body}). Notice that
using a smaller range would result in more particles, 
and therefore more accurate deformations. However, this 
would impact performance by causing worse running-times 
during the deformation algorithm, as shown in 
Table~\ref{tbl:soft_body_running_times}.

When a user applies a force to a particle, its position changes 
and this movement affects the position of all vertices 
of the particle. The particle's velocity is 
changed, proportionally to the displacement, always pointing 
to its initial position (simulating elasticity).
For skinned models, the velocity and displacement of the 
particles are calculated based on the pose of the model 
at the existing time, similar to how the skinning 
algorithm works for the vertices. The physics part is 
natively handled by the game engine employed. 

In order to apply soft-bodies clustering 
for cut, teared or drilled objects, we update the 
clustering map by adding or removing vertices, depending 
on the case. Simple rules allow fast updates without
the need to add more particles. An example of such a rule is 
that, vertices belonging to opposite sides of a tear, 
although close enough, cannot belong to the same particle, as
this would result to non-physically correct deformation results.

\subsection{Optimization Techniques}\label{sub:tools}

To perform either of the three main algorithms 
(cut, tear or drill) in near real-time, we have 
created data structures containing, mainly, vertex neighbours and 
face neighbours. 
Currently, we maintain two adjacency data structures, tailored to 
optimize the running times of the tear and the drill algorithm 
respectively. 
When importing the model, we initially check and remove any
duplicate vertices that may occur as they cause artifacts in 
our algorithms. 
Then the data structures are populated; this is the most 
time-consuming part of the preprocessing. 
After the initialization is complete, the data structures are 
updated every time the user performs one of the algorithms. 
The time required to update them is miniscule compared to the initialization 
time and is a fraction of the time required to perform the algorithm 
itself. 
A summary of the running times for the preprocessing and the main 
algorithms can be found in  Table~\ref{tbl:ctd_running_times}.

The visualization quality of the drilling algorithm was also optimized.
Specifically, if one drills the model in an area and the 
number of the edges on the contour of the hole drops below a 
user-defined threshold (e.g., 10), then the drilled part 
looks more like a polygon rather than a circle. 
In this case, we split each face involved into 4 subtriangles 
by introducing new edges that connect the mid points of their edges. 
This is done recursively until the number of intersection points 
with the updated mesh are above the threshold. 
Of course, for each introduced vertex, we assign weights to 
allow subsequent model deformation.

Important under-the-hood optimizations regard the (re)design 
of the algorithms described in \cite{Kamarianakis_Papagiannakis_2021}
to take advantage of multi-threaded or parallel processing allowed 
by modern game engines such as Unity and Unreal Engine. 
Since basic geometric subpredicates such as the intersection of a 
face of the mesh with a plane (used multiple times when cutting or tearing)
can be down in parallel, we have carefully structured our algorithms 
to achieve minimum running times. The updates of the data structures 
are also down in parallel, after each algorithm call. 

\section{Results}\label{sec:results}

The results of our methods are depicted in 
Figure~\ref{fig:teaser} and are further illustrated
in the video accompanying this work. .
% , \ref{fig:rotated} 
% and \ref{fig:soft_body}. 
We depict a cactus model of 15819 vertices and 25267 triangles
that is getting cut, teared and/or drilled. 
In all cases, further soft-body deformations and/or model 
animation are possible.
Our methods are also partially implemented in a VR medical 
training application \cite{CVRSB2021}, running on a modern
game engine.

Table~\ref{tbl:ctd_running_times} contains the time required to 
perform the model preprocessing and individually apply one of 
the algorithms for cut, tear or drill. 
In Table~\ref{tbl:soft_body_running_times}, 
we describe the time required to evaluate the soft body  
deformation depending on the number of the particles. 
All running times were obtained using an Intel core i7 7700HQ 
at 2.8GHZ with an Nvidia GTX 1050ti m (8GB RAM) graphics card.

\begin{table}[tbp]
  \caption{Running times regarding the cactus model.
  The model optimization time corresponds to the time 
  we need to detect and delete duplicate vertices of the mesh. 
  The preprocessing times required for the tear, drill and 
  clustering are only 
  performed during the initial import and can be done in parallel in 
  different threads. The running times of cut, tear and drill grow 
  linearly with respect to the number of intersection points and 
  include the time required to update the data structure that was 
  initialized during the model preprocess.  }
  \begin{center}
  \begin{tabular}{|c|c|c|}
  \hline
  Procedure & Running time  \\
  \hline
  \hline
  Model Optimization &  0.005 sec\\
  \hline 
  Preprocess for Tear & 13.620 sec\\
  Preprocess for Drill & 0.876 sec\\
  Soft Body Clustering &  41.275 sec\\
  \hline 
  \multirow{2}{*}{Cut} & 0.440 sec \\
  & (156 intersection points) \\
  \hline 
  \multirow{2}{*}{Tear} & 0.095 sec \\ 
  & (26 intersection points)\\
  \hline 
  \multirow{2}{*}{Drill} & 0.201 sec \\ %UpdateNeighs time: 0.080s
  & (27 intersection points) \\
  \hline 
  \end{tabular}
  \label{tbl:ctd_running_times}
  \end{center}
\end{table}

\begin{table}[tbp]
  \caption{Running times for the soft body deformation for the cactus 
  model. As expected, the running times are proportional to 
  the number of the particles.}
  \begin{center}
  \begin{tabular}{|c|c|}
  \hline
  Particles number & Running time min-max \\
  \hline
  \hline
  224 & 1.5 msec - 3.7 msec\\
  \hline
  452 & 1.6 msec - 4.2 msec\\
  \hline
  863 & 2.0 msec - 6.7 msec\\
  \hline 
  \end{tabular}
  \label{tbl:soft_body_running_times}
  \end{center}
\end{table}

\section{Conclusions \& Future Work}\label{sec:conclusions_future_work}

We have presented an algorithm that allows a user to perform 
unconstrained cuts, tears and drills on a rigged model in VR, while 
preserving its ability to be deformed as a soft-body. 
Since our method is geometry-based, it does not require 
significant GPU/CPU resources, it 
is amenable to work in real-time VR for even low-spec devices, making 
it ideal for mobile VR. We expect that 
it will eventually pave the way to alter the modern landscape of such VR interactions, where similar operations are mostly predefined and most state-of-the-art expensive, physically-correct methods (e.g. Finite Element Methods) cannot be used as they require significant computing resources and/or produce low fps results unsuitable for mobile VR applications.
In an area of the 3D model that the surface triangles are larger in comparison to the drilling diameter and not particularly dense, the resulting hole looks more like a polygon than a drilled circle. That's due to the low number of the intersection points of the circle with the triangles in the drilling area. In the current state of our drilling algorithm, when we increase the model density the texture coordinates are showing some artifacts due to non-proper UV recalculation. In the future, we will be addressing this artifact.

In the future, we also intend to optimize further our 'deep-cut' framework so that they can achieve higher frame-rates with compute and geometry shaders for the pre-processing stage; so far, we have minimum latency after applying any algorithm that in some cases is negligible as the user usually takes some time for mental preparation between actions. However, we are planning to support consecutive tears, something that is not possible at this stage.

As cuts, tears and drills are especially useful for VR medical training 
scenarios, we would like to explore how our framework could adapt to 
the collaborative needs of such applications. Lastly, 
we would like to see how deep learning could be used to help
identify which type of clustering is best suited, based on the model
and the action(s) the user would like to perform.

\begin{acks}\label{sec:acks}
This project was co‐financed by European Regional Development 
Fund of the European Union and Greek national funds through the 
Operational Program Competitiveness, Entrepreneurship and Innovation, 
under the call RESEARCH – CREATE - INNOVATE 
(project codes:T1EDK-01149 and T1EDK-01448). 
The project also received funding from the European Union’s 
Horizon 2020 research and innovation programme under grant agreements 
No 871793(ACCORDION) and No 101016509(CHARITY).
\end{acks}

\bibliographystyle{ACM-Reference-Format}
\bibliography{bibliography}

\end{document}